\documentclass[preprintnumbers,twocolumn,preprintnumbers,amsmath,amssymb,nofootinbib]{revtex4}

\usepackage{graphicx}% Include figure files
\usepackage{epsfig}
\usepackage{color}

\newcommand{\beq}{\begin{equation}}
\newcommand{\eeq}{\end{equation}}
\newcommand{\bea}{\begin{eqnarray}}
\newcommand{\eea}{\end{eqnarray}}
\newcommand{\bwd}{\begin{widetext}}
\newcommand{\ewd}{\end{widetext}}

\begin{document}

\title{A proposal to fight tornadoes with multiple connected balloons}

\author{Ji Qiang}
\email{jqiang@lbl.gov}
\affiliation{...}
%Lawrence Berkeley National Laboratory,
%          Berkeley, CA 94720}

\begin{abstract}
A tornado is an extreme weather condition that can cause enormous damage to human
society. In this paper, we propose a low cost, environmentally friendly method to 
fight against tornadoes using clusters of connected balloons. 
Deployed over a sufficiently wide area, these large balloons may be able to
reduce the wind speed of the tornado, block and disrupt the convective flow
of air, and destroy the tornado.
\end{abstract}
%1)should I add some discussions about this method is
%to fight the tornadoes but not the tornado formation??
%2)should I mention about the effects of surface roughness on the
%tornadoes??
%This method has two impact:
%1)block the updraft, the downdraft air flow
%2)divert the convective air flow.

\maketitle

\section{Introduction}

%\section{Past tornado mitigation proposals}
A tornado is an extremely dangerous weather condition with a violently rotating column of air extending 
from the base of a thunderstorm down to the ground~\cite{wiki,safety}. 
Through human history, tornadoes have caused many human fatalities, 
injuries, and massive economic damage throughout the world. For example, on April 26, 1989, a tornado in Bangladesh
killed about $1,300$ people~\cite{bang}. 
The United States has the most tornadoes in the world, nearly four times more than those estimated in all of Europe.
In the United States, there are about 1000 tornadoes each year 
causing about on average 90 fatalities per year in the past 10 years from 2009 to 2018~\cite{damage},
more than 1000 injuries per year,  
and %\$850
several hundred million dollars worth of damage per year~\cite{loss}.
%The United States has the most tornadoes in the world, nearly four times more than estimated in all of Europe.
%excluding waterspouts.[
%his unique topography allows for frequent collisions of warm and cold air, the conditions that breed strong, long-lived storms throughout the year
%
%, There are especially large number of tornadoes during
%the spring and summer seasons in the tornado alley of the State. 

In atmospheric sciences, extensive research was done to understand 
dynamics and genesis of tornadoes~\cite{ward,church,markowski2003,yih,doswell2007,makarieva,rotunno,naylor,markowski2014,jones,amots,baker,coffer,refan,romanic}.
However, there were few studies about
mitigation of tornadoes given the challenge of the problem.
An old idea that was frequently mentioned is to use nuclear weapons to bomb the tornado~\cite{faq}. This idea is based on 
the assumption that the large energy released by a nuclear weapon explosion above a storm would heat the cold air there 
and disrupt the storm.
This idea was dismissed since it would also cause severe damage to surrounding environment, 
bring about negative effects on human health, and even destroy private properties.
Another idea to destroy tornadoes is to use microwave beams 
from a number of solar-powered space satellites~\cite{microwave}.
This idea is based on the assumption that the formation of a tornado needs cold downdraft
air. If one can heat the cold downdraft air, this would prevent the tornado from forming.
To heat the cold downdraft air, this idea is to use a number of satellites in space to
collect solar energy, convert it into microwaves,
and send the microwave beams down to Earth to heat the cold downdraft air flow.
However, this method would be extremely expensive and was refuted in reference~\cite{doswell0}.
Recently, three man-made great walls were proposed in references~\cite{tao1,tao2}
to prevent the formation of tornadoes in the US Tornado Alley. 
The assumption behind this idea is to use the man-made wall to reduce the wind
speed, weaken air mass collisions, and eliminate the major tornadoes in this area. 
The author proposed  
building three great walls, each with 300 m height and 50 m width, in this area.
The first one is near the northern boundary
of the Tornado Alley, maybe in North Dakota. The second one is in the
middle, maybe in the middle of Oklahoma and going to east. The third one is
in the south of Texas and Louisiana.
However, building those great walls or skyscrapers would be a very expensive project by itself
and the idea was criticised by the other meteorologists~\cite{dahl}.
Another idea to prevent the tornadoes is based on the assumption to change the local climate
by cloud seeding. However, there was evidence suggesting that cloud
seeding might not help prevent formation of tornadoes, but increase formation
of tornadoes in a number of cases~\cite{doswell}.
So far, none of the above methods are environmentally and economically practical 
to mitigate tornadoes.
%prevent
%the formation of tornadoes.

A tornado is normally associated with the turbulent instability of 
atmospheric air flow in a storm.
In physics, in order to suppress a coherent instability, one can 
introduce a method to randomize the air flow and to enhance the
incoherent motion of atmosphere.
In this paper, we propose a low cost, environmentally friendly method to fight tornadoes by
using multiple connected large balloons.
Each balloon can have a size of about ten meters. A number of these
large balloons are 
connected together to form a cluster that can
have a few ten meters across and can cover an area of thousands of square meters.
Multiple clusters of these balloons flying into a tornado would cause decoherent motion
of air, disrupt the high speed rotating wind, block
the warm updraft air flow, and potentially destroy the tornado.
Furthermore, these balloons can be reused multiple times for different tornado seasons 
and can be deflated and stored in a small building when a tornado season is over. 
There is no modification of the local climate or any other environmental factors.

The organization of this paper is as follows: after the Introduction, we briefly review
the life cycle of a tornado in Section II and characteristics of tornadoes in Section III; we then
present the proposed multiple connected balloon method in Section IV; discuss potential challenges 
in Section V; and give some further discussions in Section VI.
%other applications in Section VI.

\section{Life cycle of a tornado}
A tornado is a dangerous and complex weather phenomenon.
Most violent tornadoes are generated from the supercell storm.
A thunderstorm can form when the warm humid air collides with the cold air.
A supercell is a severe thunderstorm that contains mid-level persistent
updrfat rotation (also called mesocyclone).
Even though it is still not fully understood how exactly tornadoes form, grow,
and stop,
some basic understanding of a tornado formation can be illustrated using three steps
in Fig.~1~\cite{markowski2014}.
\begin{figure}
\includegraphics[angle=0,width=.40\textwidth]{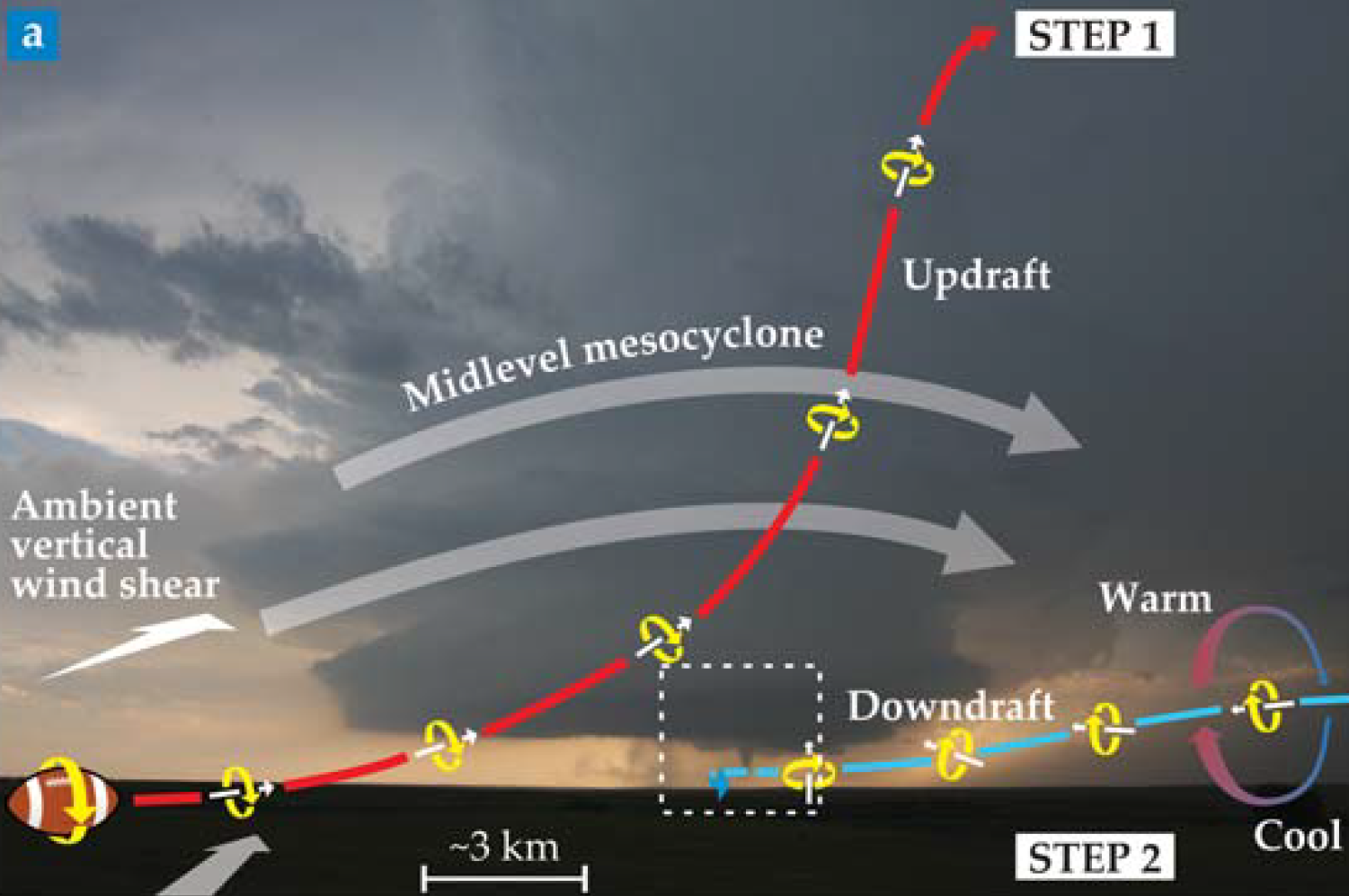}
\includegraphics[angle=0,width=.28\textwidth]{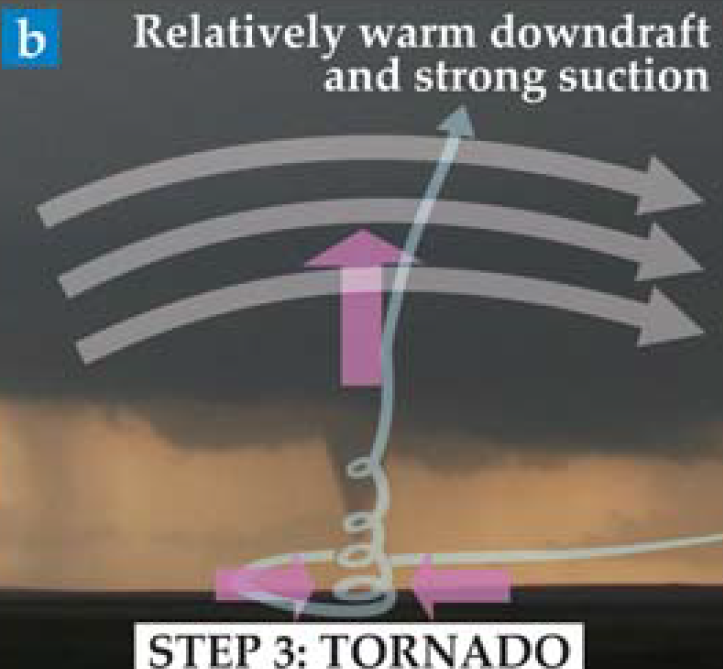}
\caption{An illustration of tornado formation from a supercell storm following the three 
steps in reference~\cite{markowski2014}.}
\label{fig1}
\end{figure}

%There are three steps in the formation of a tornado from a 
%supercell thunderstorm as shown in Fig.~1.
In step one, the vertical wind shear of the air flow will generate horizontal rolling of air, i.e. horizontal vorticity.
Here, the vertical wind shear denotes the change of horizontal wind speed and
direction along the vertical altitude.
The horizontal vorticity is tilted upward by the warm updraft wind 
field at the mid-level altitude to become vertical vorticity,
i.e. mesocyclone. 
In step two, the precipitation in the thunderstorm 
drags with it a region of quickly descending cold air to form 
the downdraft air flow. This downdraft flow can coexist with the updraft flow since the 
precipitation normally falls outside the 
updraft region. This causes a buoyancy gradient at low altitude and generates horizontal vorticity near the ground. These horizontal
vorticity is tilted upward by the surrounding wind field to form near ground vertical vorticity as shown in step two of the Fig.1.
The near ground rotation by itself will not develop into a strong tornado. 
In step three, in the region dominated by rotation, the pressure is lower than the outside region. 
This produces an upward-directed pressure gradient force that sucks the surrounding air into the region.
This dynamic updraft suction intensifies the ground level rotation through the conservation of angular momentum and a violent high rotating speed tornado is formed.

Initially, the tornado has a good source of warm, moist air flowing inward to power it. 
Without surface friction, it grows until the
centrifugal force balance the inward pressure-gradient force. 
As the cold downdraft air flow completely wraps around and cuts off the tornado's air supply,
the updraft begins to weaken, the tornado evolves into the dissipating stage.
This stage often lasts no more than a few minutes until the tornado ends.

\section{Characteristics of tornadoes}

Tornadoes are visualized with strong rotating wind and a cloud of debris. 
Most tornadoes have a narrow funnel shape except that in the dissipating stage, the tornadoes can resemble narrow tubes or ropes, or
can twist into more complex shapes. The size of tornadoes can vary widely from case to case. Some weak or strong dissipating
tornadoes can be very narrow and only a few meters across. Some can be as wide as a mile or more. Most tornadoes are about
80 meters across while on average, tornadoes are about 150 m across~\cite{wiki}.

Most tornadoes have a rotating wind speed less
than 110 miles per hour while some extreme tornadoes can have a speed over 300 miles per hour.
Tornadoes are classified into six categories using intensity Enhanced Fujita (EF) scale, EF-0 through EF-5, according to their wind speeds and the damages~\cite{wiki2}. 
EF-0 tornadoes are the mildest with wind speed between 65 and 85 Miles Per Hour (MPH), causing light damage;
EF-1 have wind speed between 86 and 110 MPH, causing moderate damage; EF-2 wind speed between 111-135 MPH, causing
considerable damage, EF-3 wind speed between 136-165 MPH, causing severe damage; EF-4 wind speed between 166-200 MPH, causing
devastating damage; EF-5 wind speed greater than 200 MPH, causing incredible damage.
F-5 tornadoes are the most dangerous with wind speed between 261 and 318 MPH and cause violent damage to surroundings.

Tornadoes can last from several seconds to more than an hour~\cite{faq}.
Most tornadoes last less than 10 minutes and move at speeds of about 10 to 20 miles per hour. 
The average distance tornadoes have traveled is about three and half miles.
In addition to strong winds, tornadoes also show changes of temperature, moisture, and pressure of surrounding air. 
Temperature tends to decrease and moisture content to increase around a tornado.
From the outside of the tornado to the center of the tornado, there could be about
100 mbar pressure difference~\cite{eric}.
Accompanying the strong wind of tornadoes, there can be very heavy rain, frequent lightning, and hail in the storms.

\section{Mitigation of tornadoes using multiple connected balloons}

%\begin{figure*}
\begin{figure}
\includegraphics[angle=0,width=.30\textwidth]{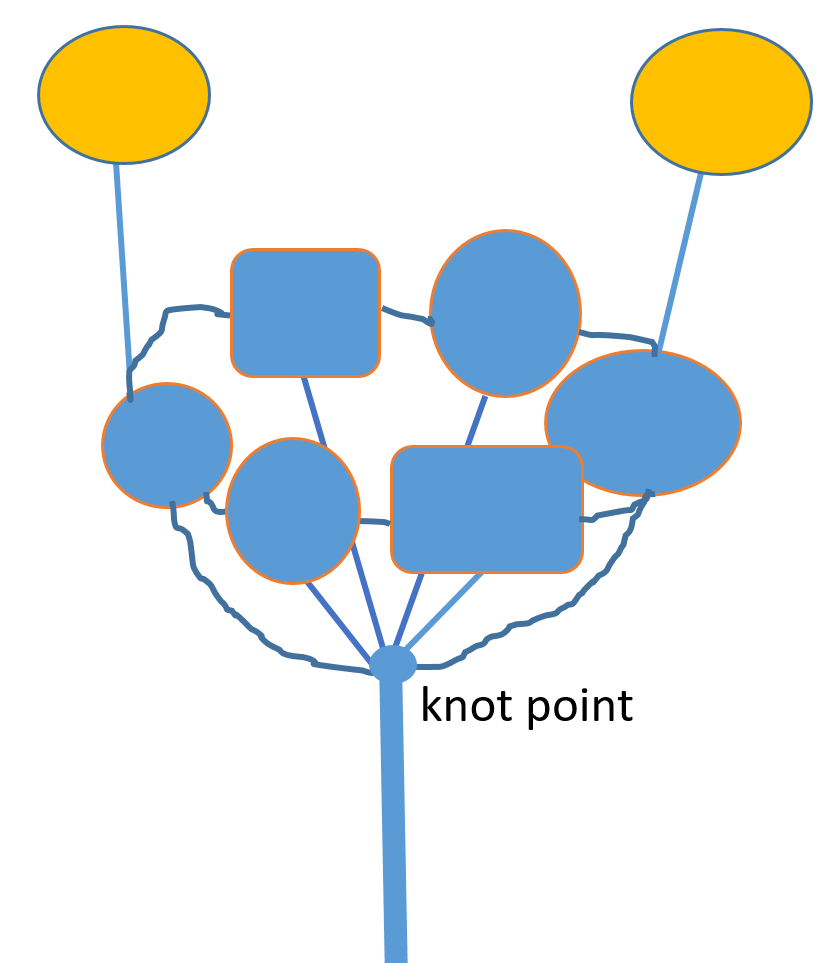}
	\caption{A schematic plot of a cluster of multiple connected balloons.
	Here, the balloons inside the cluster can have different sizes and shapes to
	enhance the random irregular motion of air. Two extra balloons (yellow) filled with
	light gas such as helium are used as lifting balloons.}
\label{fig2}
\end{figure}
%\end{figure*}
From the life cycle of a tornado, we know that the formation and the survival of the tornado depends
on the downdraft and the updraft air flow. Without the updraft warm air flowing into the tornado, 
it would
become weak and dissipate quickly. From the characteristics of tornadoes, we know that even though
the area affected by a thunderstorm can be quite large ($\sim$miles), the size of a typical tornado is relatively
small ($\sim$100 meters). This suggests that if one could block the updraft warm air 
flowing into a 
tornado, the tornado might be stopped quickly.

In this section, we discuss fighting tornadoes using multiple connected balloons.
This method is to use multiple connected balloon clusters to block the updraft and the swirl
air flow of a tornado and to divert the convective air flow into the disordered air flow
to achieve the goal of slowing down and even destroying the tornado.
A schematic plot of a connected balloon cluster is shown in Fig.~\ref{fig2}.
In this cluster, multiple balloons (six balloons in this example), 
each with individual string (this string
by itself can consist of multiple strings), are tied together with roughly the same distance
from the balloon to the knot point to form a star like structure. Then those balloons
are connected side by side with a string to form a polygon like structure. Using the string
connection between balloons is to enhance the random interaction among balloons and to
cause the disordered motion of air. Extra balloons (two in this example), each connecting 
to one of these six balloons, 
are used to help lift the balloon cluster. The different sizes and shapes (round, rectangular box, 
and ellipsoid) of balloons are included there to illustrate that those balloons can
have irregular shapes instead of just the round shape. This could even enhance the random motion of
balloons after collisions, helps disrupt the coherent convective
air flow, slows down the wind speed of the tornado. 
These balloons also block the inflow warm updraft air from the tornado suction. 
This leads to the dissipation of the tornado and eventually destroys the tornado.

%\begin{figure*}
\begin{figure}
\includegraphics[angle=0,width=.20\textwidth]{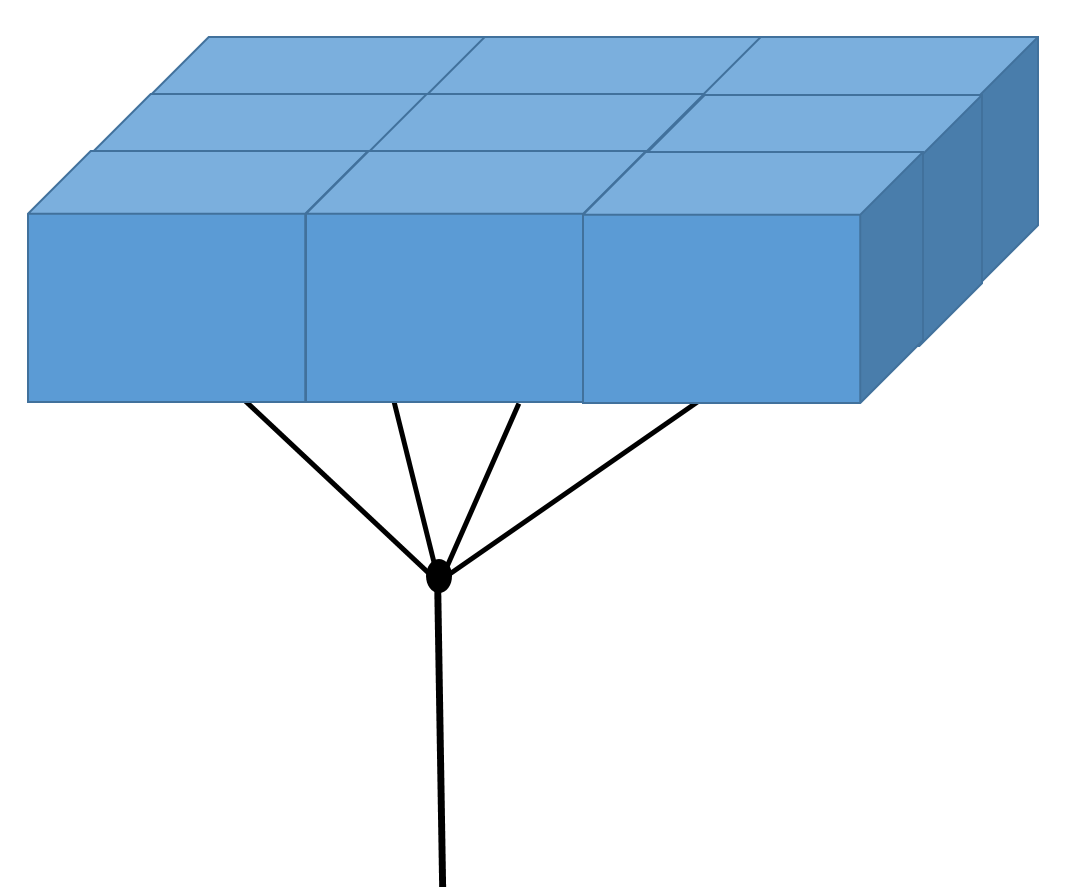}
\caption{A schematic plot of a single multi-chamber
	rectangular shape balloon.}
\label{fig3}
\end{figure}
%\end{figure*}
Figure~3 shows a rectangular shape balloon that consists of
multiple chambers.
There are total nine chambers in this example. 
Using multiple chambers in a balloon helps it to float inside
the tornado after being
penetrated by high speed flying debris.
It also reduces the time to pump 
the balloon through multiple inlets. 
The size of the balloon can be on the
order of ten meters in the two horizontal dimensions and on the order of meters 
in the vertical dimension. 
Such a rectangular shape balloon might be more efficient to 
block the updraft air flow in the tornado than the round shape balloon.
The length of the string connecting two balloons in a cluster can be on the order
of 20 to 30 meters. The string from the balloon to the knot point of the cluster 
can have similar length.
A cluster of these connected balloons has an across of 
about 20 to 50 meters and covers an area of thousands square meters. 
A dozen of clusters will cover the size of a football field, which
is about the size of a typical tornado.
Each balloon can be made of materials such as air bed vinyl with a rough soft surface.
The rough soft surface helps slow down the wind speed at the surface of the balloon.
The balloon can be electrically pumped with
air or with helium or with cheaper hydrogen (if fire is not a problem). 
The lifting balloons can
be filled with helium or with hydrogen so that they will get into the tornado storm
first and
drag the entire cluster into the tornado with the help of updraft suction force of 
the tornado.
Each balloon consists of multiple independent air chambers so that even one is broken,
the balloon can still fly. The altitude of these balloons in the tornado will be relatively 
low (e.g. maybe less than 30 meters) and can be made lower or higher by adjusting
the length of the large rope connecting the cluster using 
a remote controller. Varying the altitude of the balloon cluster inside a tonorado might
help enhance the random motion of air flow inside the tornado.
%controlled, automatically adjustable large rope as shown in
%Fig.~2. 
As the tornado moves, the large nylon rope is released and becomes longer
so that the balloon cluster can move together with the tornado (like fishing). 
%Depending on the size
%of the tornado, a number of those balloon clusters can be used to fight against the
%tornado.

%\begin{figure*}
\begin{figure}
\includegraphics[angle=0,width=.40\textwidth]{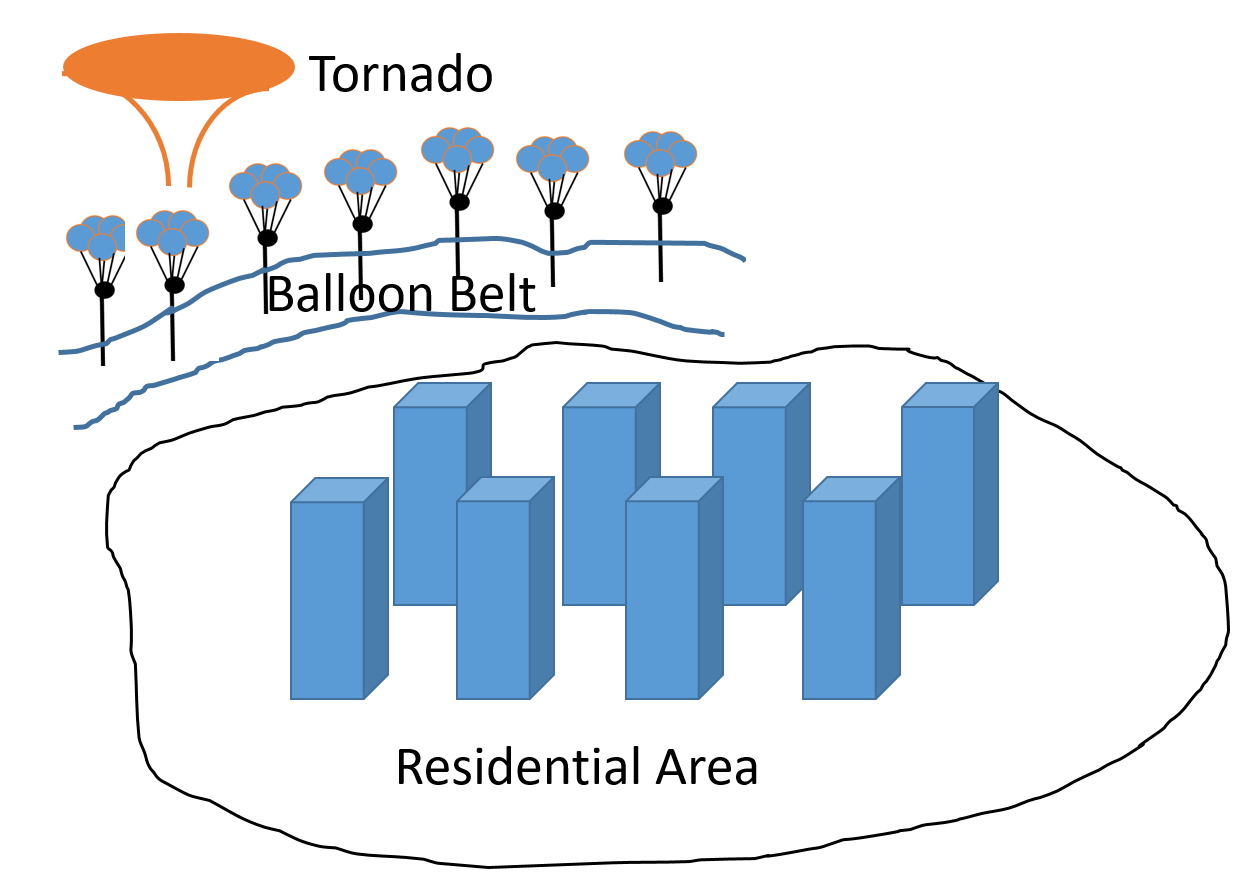}
\caption{A schematic plot of deploying a balloon belt
near a residential area to fight against the tornado.}
\label{fig4}
\end{figure}
%\end{figure*}
Figure~4 shows a schematic plot of the deployment of these balloon clusters in front
of a residential area. An open space near the residential area can be used to deploy
those balloon clusters. 
Using an open space has the advantage of reducing the amount
of debris in a tornado. 
A number of electric pumping spots and cluster rope holders can be
installed in the open space. 
On an ordinary day without tornadoes, the area
is open to the general public. On a tornado day, 
 multiple clusters of balloons will be quickly pumped and released into the air to
 form a balloon belt. 
 The width of the balloon belt can be about one hundred meters
 and the length of the belt can be a couple of hundred meters.
 Once a tornado moves into the balloon belt, it will interact with the balloon clusters
 inside the belt. These clusters block and disorder the convective air flow in the tornado
through the random motion of balloons and cause the tornado to lose its 
high wind speed before reaching the residential area.
This would protect the residential area from the damage of strong wind
and flying debris of the tornado. 
If possible, these balloon clusters might also be installed on a strong mobile 
device (e.g. a tank-like low weight center heavy automobile)
and be carried directly into the path of the tornado to fight against the tornado. 

\section{Potential challenges of the proposed method}
There are some potential challenging questions for the above proposed method.

{\it Firstly, are these balloons large enough to affect a tornado?}
As mentioned in Section III, the size of a storm that helps form the tornado can be large
(on the order of miles), but the size of a tornado that touches the ground can
be much smaller. A typical tornado has a width of about hundred meters and a size of
about a football field. Using multiple clusters of balloons might not be able to
affect the local weather. However, 
%given the size of a balloon cluster,
a number of clusters that contain a few hundred large balloons will cover
an area greater than a football field and 
should be able to affect the local air flow 
inside and around the tornado.
Some past studies of debris loading in a tornado suggested
that under some conditions, the debris loading could have 
substantial impact on the tornado dynamics, causing $50\%$ or more reduction of near-surface
wind speeds~\cite{lewellen,bodine}. 
Given the much larger size of the balloon compared with the debris, the impact of these balloons on
the tornado dynamics should be more significant.

{\it Secondly, will the balloons be strong enough to survive the strong wind of a tornado?}
The strong wind of the tornado can break down trees and destroy houses. It might also
break a balloon. However, the difference between the flying balloon and
a tree or a house is that the latter is still while the former moves
together with a wind. The relative speed between the wind and the tree or the house is 
large for a strong wind. On the other hand, the relative speed between the strong wind and
the moving balloon might not be very large since the balloon moves and rotates together with
the wind. This reduces the impact of the strong wind on the balloon.
Each balloon can be tied by multiple nylon strings at several locations.
These strings are braided together to form a single nylon string as
shown in Fig.~3. In case that a single string is broken,
the other strings can still hold the balloon. Such a balloon is further
bundled and connected with the other balloons to form a balloon cluster.
This cluster will rotate with the swirl wind inside the tornado. 
The balloons inside the cluster interact with each other due to 
different rotating speeds.
These moving balloons would disrupt the convective air flow, 
reduce the rotating speed of air inside the tornado, and
block the updraft air flowing into the tornado.

{\it Thirdly, will the flying debris break the balloon?}
The flying debris inside the strong wind of a tornado can have a very high speed 
and a large momentum. It would penetrate into or through the balloon. However, this 
would not prevent the balloon from flying since the balloon is driven
by both the updraft inflow wind from the suction of the tornado and the dragging
force of the lifting balloons. 
Also, the balloon consists of multiple independent
chambers. If one chamber is broken, the other chambers can still function.
Moreover, the balloon can be made of air bed like material or multi-layer material (including a special layer). 
Even after the balloon is pierced through by a flying object, 
it would still hold sufficient air to be able to float in the strong wind (like a kite).
On a rainy storm day, the soft surface of the balloon can soak water and 
increase the balloon weight. The heavier weight of the balloon helps dissipate more
convective wind energy of the tornado. 
%The balloon can be made of multiple layers.
%If the balloon is penetrated by debris, most air inside the balloon can still stay.
%Another possible way to avoid the debris is to fly the balloon into higher altitude
%before being sucked into the tornado.

{\it Fourthly, will the pressure drop inside a tornado blow up the balloon?}
The pressure drop from the region outside of the tornado to the region inside
the tornado is about $100$ mbar. This corresponds to about $10\%$ decrease of 
the usual atmosphere pressure. A balloon made of the air-bed material or 
other type of materials
should be able to withstand this pressure difference.

{\it Fifthly, what if a tornado moves through these balloons?}
The tornado will move with a relatively slow speed (a typical speed is about 
ten to twenty miles per hour). The length of the nylon rope connecting to
the balloon cluster can be made 
automatically adjustable (with a remote controller) so that the cluster 
moves together with the tornado. This is similar to fishing. 
Once the balloon cluster is caught by the tornado, it would move with the tornado until
the end of the tornado.
%the random motion the connected balloons dissipates the convective
%air flow inside the tornado.
%the coherent vortex air inside the tornado

%\section{Discussions}
\section{Further discussions}

After a tornado storm is over, balloons hit by flying debris
can be repaired and reused for another tornado. 
The cost of a balloon can be a few hundred US dollars. The cost of a cluster of 
connected balloons can be a few thousand dollars. A dozen of clusters of
balloons would cost on the order of ten thousand dollars. Compared with 
the multiple million dollar damage and even life lost caused by the tornado, 
the cost of these balloons is much less. Also, these balloons would not affect
the local weather condition, neither cause any environmental problem. After the
tornado storm, these balloons can be deflated electrically and folded into a small
volume and stored in a convenient place. On a tornado day,
these balloons can be quickly pumped using electric devices, set up, and released
into the air within a couple of minutes. Given the current warning time of the tornado (more
than ten minutes)
it should be sufficient to set up the balloon system within the warning time.
In practice, it will be safer to set up the balloon system somewhat earlier so that
people can still have time to find a safe shelter after that.

\begin{figure}
\includegraphics[angle=0,width=.40\textwidth]{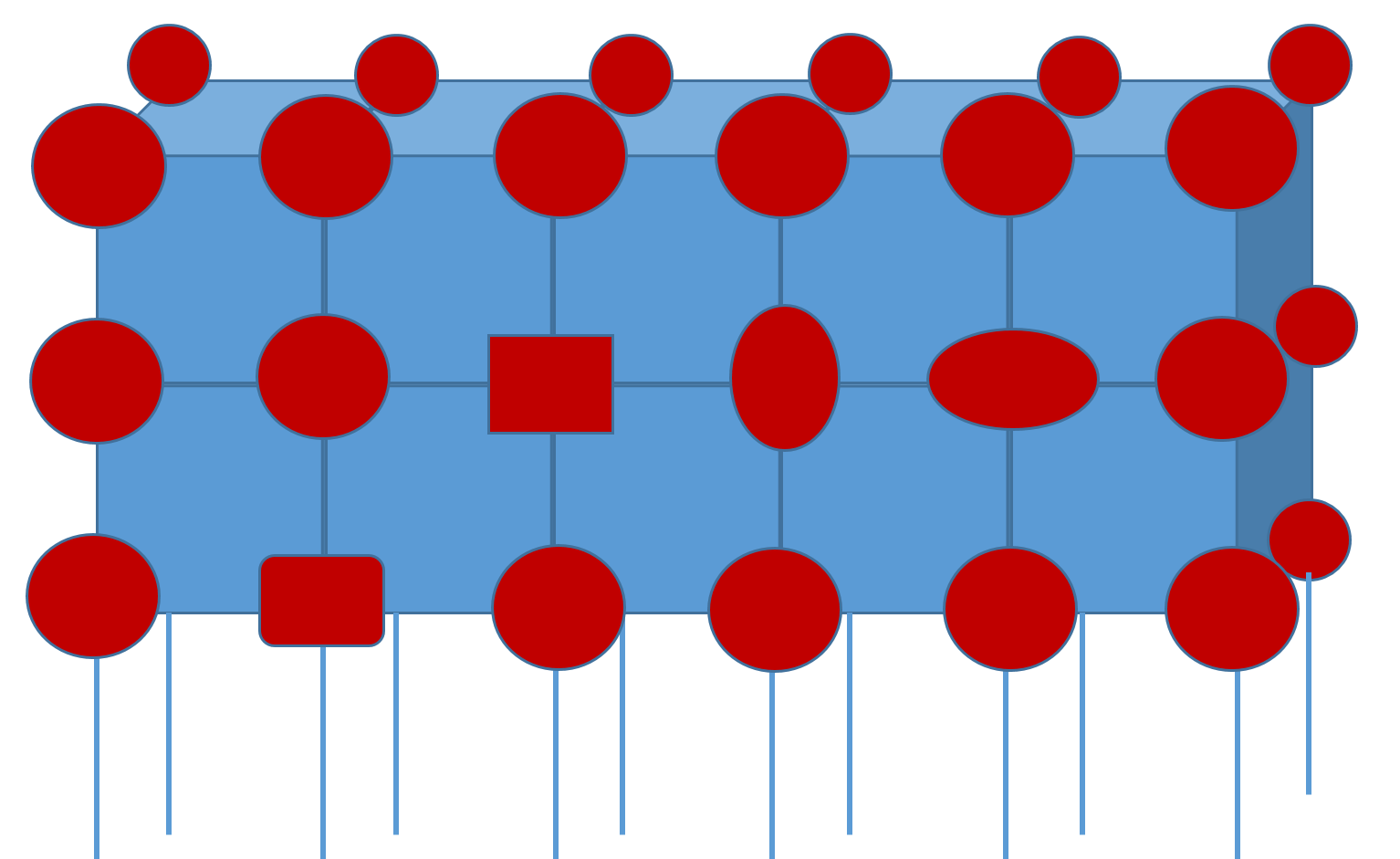}
\caption{A schematic plot of a multiple connected balloon wall.}
\label{mesh}
\end{figure}

Besides the star like cluster structure, these large size balloons can also be 
connected into a mesh like structure
to form a protecting balloon wall. A schematic plot of such a balloon wall
is shown in Fig.~\ref{mesh}. Again, the shape and the size of individual balloon
can be different from each other. Such a wall in front of a residential area
helps slow down the strong wind blowing from a severe storm and protects the people 
and properties behind it. Such a balloon wall can also be used test the great
wall idea proposed in references~\cite{tao1,tao2} with a much lower cost. 
For a 300 meter high and 50 meter wide wall, it will use about 100 
($25\times 4$) ten meter size balloons. Assuming that the cost of each
balloon is between $500$ and $1000$ US dollars, the cost of such a great
balloon wall will be between $50,000$ to $100,000$ US dollars. This
is much cheaper than the similar size skyscraper. Even one uses
a two-plane, three-dimensional structure as shown in Fig.~\ref{mesh} by
doubling the number of balloons in the wall, it is still much
less expensive than a skyscraper great wall.

In this paper, we propose a method to use multiple connected balloons to fight
against tornadoes. This method is environmentally friendly and economically practical.
However, a number of further studies probably need to be done in real applications.
These studies include the optimal choice of size and shape of the balloons, of the length
of string connecting the balloons, of the number of balloons in a cluster, of the material
used to make the balloons, of the weight of the balloons, of the gas used to fill the balloons, of the altitude
of the balloons, and of the way to deploy the balloons. 
This method can be tested 
with the experimentally simulated tornadoes and if possible with the numerically 
simulated tornadoes. 

%Another possible alternative is to use flying balloons to drop parachutes.
%Those parachutes will help lift the lower layer balloons and also
%block the downdraft and the updraft air flow inside the tornado.
%The advantage of using parachutes is to 
%One can also use wind kite to lift the lower layer balloons.

%\section{ACKNOWLEDGEMENTS}
%This work was supported by the U.S. Department of Energy under Contract No. DE-AC02-05CH11231, and
%used computer resources at the National Energy Research
%Scientific Computing Center.


\begin{thebibliography}{99}
\bibitem{safety}https://www.weather.gov/safety/tornado.
\bibitem{wiki}https://en.wikipedia.org/wiki/Tornado.
\bibitem{bang}https://en.wikipedia.org/wiki/Daulatpur\\\%E2\%80\%93Saturia\_tornado
\bibitem{damage}https://www.weather.gov/hazstat/ 
\bibitem{loss}https://sciencing.com/causes-effects-tornadoes-8204458.html
\bibitem{ward}N. B. Ward, "The exploration of Certain Features of Tornado Dynamics using a laboratory model," J. Atmospheric Sciences 29, p. 1194, 1972.
\bibitem{church}C. Church, D. Burgess, C. Doswell, R. Davies-Jones, ed.``The Tornado: Its Structure, Dynamics, Prediction, and Hazards,'' 
	Geophysical Monograph Series, American GeoPhysical Union, 
Published under the aegis of the AGU Books Board, 1993.
\bibitem{markowski2003}P. Markowski, J. M. Straka, E. N. Rasmussen, ``Tornadogenesis resulting from the transport of circulation by a downdraft: idealized numerical simulations,'' J.  Atmospheric Sciences, vol. 60, p. 795 (2003). 
\bibitem{yih}C. Yih, ``Tornado-like flows,'' Physics of Fluids 19, 076601, (2007).  
\bibitem{doswell2007}C. A. Doswell III, ``Historical overview of severe convective storms research,'' Electronic J. Severe
Storms Meteor., 2(1), 1-25.
%\bibitem{beatty}Beatty, K., E. N. Rasmussen, and J. M. Straka, 2008: The supercell spectrum. Part I: A review of research
%related to supercell precipitation morphology. Electronic J. Severe Storms Meteor., 3 (4), 1-21.
\bibitem{makarieva}A.M. Makarieva, V.G. Gorshkov, ``Condensation-induced kinematics and dynamics of cyclones, hurricanes
and tornadoes,'' Physics Letters A 373 (2009) 4201-4205.
\bibitem{rotunno}R. Rotunno, ``The Fluid dynamics
of tornadoes,'' Annu. Rev. Fluid Mech. 2013. 45:59-84.
\bibitem{naylor}J. Naylor and M. S. Gilmore, ``Vorticity Evolution Leading to Tornadogenesis and Tornadogenesis
Failure in Simulated Supercells,'' J. Atmospheric Sciences 71, p. 1201, 2013.
\bibitem{markowski2014}P. Markowski, and Y. Richardson, ``What we know and don't know about tornado formation,'' 
	Physics Today 67, 9, p. 26 (2014).
\bibitem{jones}R. Davies-Jones, ``A review of supercell and tornado dynamics,'' Atmospheric Research 158-159 (2015) p.274.
\bibitem{amots}N. Ben-Amots, ``Dynamics and thermodynamics of a tornado: Rotation effects,'' Atmospheric Research 178-179 (2016) p. 320.
\bibitem{baker}C.J. Baker, M. Sterling, ``Modelling wind fields and debris flight in tornadoes,'' Journal of Wind Engineering \& Industrial Aerodynamics 168 (2017) p.312.
\bibitem{coffer}B. E. Coffer and M. D. Parker, J. M. L. Dahl, L. J. W. and A. J. Clark, 
	``Volatility of Tornadogenesis: An Ensemble of Simulated Nontornadic
and Tornadic Supercells in VORTEX2 Environments,'' Monthly Weather Review 145, p. 4605, 2017.
\bibitem{refan}M. Refan, H. Hangan, ``Near surface experimental exploration of tornado vortices,'' Journal of Wind Engineering \& Industrial Aerodynamics 175 (2018) p.120.
\bibitem{romanic}D. Romanic, D. Parvu, M. Refan, H. Hangan, ``Wind and tornado climatologies and wind resource modelling for a
modern development situated in Tornado Alley,'' Renewable Energy 115 (2018) p.97. 
\bibitem{faq}https://www.spc.noaa.gov/faq/tornado/.
%\bibitem{microwave}https://www.popsci.com/scitech/article/2003-07/how-destroy-tornado/.
\bibitem{microwave}B. Eastlund and L. Jenkins, ``Taming tornadoes: Storm abatement from space,'' IEEE Aerospace Conference Proceedings 1 (2001), p. 389.
\bibitem{doswell0}http://www.flame.org/\~cdoswell/wxmod/\\
	Eastlund\_Exchanges.html.
\bibitem{tao1}R. Tao, ``Eliminating the major tornado threat in Tornado Alley,'' Int. J. Mod. Phys. B 28(22), 1450175 (2014).
\bibitem{tao2}R. Tao, ``Can we eliminate major tornadoes in Tornado Alley?-Response to the Comments'' Int. J. Mod. Phys. B 28(29), 1475005 (2014).
\bibitem{dahl}J. M. L. Dahl, P. M. Markowski, ``Comment on eliminating the major tornado threat in Tornado Alley,'' Int. J. Mod. Phys. B 28(29), 1475004 (2014).
\bibitem{doswell}C. Doswell, ``What about weather modification?,'' \\
	http://www.flame.org/\~cdoswell/wxmod/wxmod.html.
\bibitem{wiki2}https://en.wikipedia.org/wiki/Enhanced\_Fujita\_scale.
\bibitem{eric}E. R. Snodgrass, ``The science of extreme weather,'' Scholastic Library Publishing, 2016.
\bibitem{lewellen}D. C. Lewellen, B. Gong, and W. S. Lewellen, ``Effects of finescale debris on near-surface tornado dynamics,'' J. Atmospheric Sciences 65, p. 3247 (2008).
\bibitem{bodine}D. J. Bodine, T. Maruyama, R. D. Palmer, C. J. Fulton, H. B. Bluestein,
and D. C. Lewellen, ``Sensitivity of tornado dynamics to soil debris loading,''
J. Atmospheric Sciences 73, p. 2783 (2016).

\end{thebibliography}
\end{document}